\documentclass[12pt]{article}
\usepackage{times}
\usepackage{fullpage}
\usepackage[numbers]{natbib}
\usepackage[utf8]{inputenc}
\usepackage{titlesec}
\titleformat{\section}[block]{\Large\bfseries\filcenter}{\thesection}{1em}{}
\titleformat{\subsection}[block]{\large\itshape\filcenter}{\thesubsection}{1em}{}
\titleformat{\paragraph}[runin]{\itshape}{\theparagraph}{1em}{}[. ]
\usepackage{amsmath}
\usepackage{amsfonts}
\usepackage{siunitx}
\usepackage[dvipsnames]{xcolor}
\usepackage{tabu}
\usepackage{hyperref}
\usepackage{cleveref}
\usepackage{nameref}
\usepackage{graphicx}
\usepackage{lineno}
\usepackage{setspace}

\renewcommand{\r}[1]{\mathrm{#1}}

\DeclareMathOperator{\re}{Re}
\renewcommand{\d}{\mathrm{d}}
\DeclareMathOperator{\tr}{tr}
\DeclareMathOperator{\cov}{Cov}

\DeclareMathOperator{\diag}{diag}

\title{Nestedness Promotes Stability in Maximum-Entropy Bipartite Food Webs}

\author{Zhening Li\textsuperscript{1,*} \and
John Harte\textsuperscript{2}}

\date{}

\newcommand{\intro}{Introduction}
\newcommand{\flow}{A MaxEnt Energy Flow Model in a Two-Layer Network}
\newcommand{\empirical}{Empirical Testing}
\newcommand{\stability}{Nestedness Promotes Stability}
\newcommand{\discussion}{Discussion}

\begin{document}

\maketitle

\noindent{} 1. Massachusetts Institute of Technology, Cambridge, Massachusetts 02139;

\noindent{} 2. University of California, Berkeley, Berkeley, California 94720.

\noindent{} * Corresponding author; e-mail: zli11010@mit.edu.

\bigskip

\textit{Manuscript elements}: Abstract; Significance Statement; Main Text; Table 1; Figures 1 to 5; Supplemental Information.

\bigskip


\bigskip

\textit{Manuscript type}: Direct Submission 

\bigskip



\newpage{}

\section*{Abstract}

{\bf
Food web topology and energy flow rates across food web linkages can influence ecosystem properties such as stability.
Stability predictions from current models of energy flow are often sensitive to details in their formulation,
and their complexity makes it difficult to elucidate underlying mechanisms of general phenomena.
Here, within the maximum information entropy inference framework (MaxEnt),
we derive a simple formula for the energy flow carried by each linkage between two adjacent trophic layers.
Inputs to the model are the topological structure of the food web and aggregate energy fluxes entering or exiting each species node.
For ecosystems with interactions dominated by consumer-resource interactions between two trophic layers,
we construct a model of species dynamics based on the energy flow predictions from the MaxEnt model.
Mathematical analyses and simulations of the model show that a food web topology with a higher matrix dipole moment promotes stability against small perturbations in population sizes,
where the \textit{matrix dipole moment} is a simple nestedness metric that we introduce.
Since nested bipartite subnetworks arise naturally in food webs,
our result provides an explanation for the stability of natural communities.
}

\newpage{}

\section*{Significance Statement}

{\bf
Traditionally, models of population dynamics use a large number of free parameters
and are generally too complex to study analytically.
On the other hand, analytic approaches that abstract away these complexities into random variables are prone to being unrealistic.
We take an approach that strikes a balance between detail and tractability.
In a two-layer food web, we model feeding interactions in a MaxEnt framework and treat other mechanisms as random perturbations.
An analytic analysis of the model shows that a simple nestedness metric (the matrix dipole moment)
has a high positive correlation with stability, thus bridging food web topology with the stability of ecological communities.
}

\newpage{}

\section*{\intro}


Food webs are a major component of modeling population dynamics in an ecosystem \citep{dunne2009foodwebs}.
Various models and theories have been developed to predict trophic network structure
(e.g., the niche model \citep{williams2000nichemodel})
and energy flows (e.g., the allometric trophic network \citep{berlow2009atn,brose2006atn}).
Since energy flow rates determine rates of biomass change,
these models give rise to population dynamics whose stability characteristics
can be studied \citep{ruiter1995stability,berlow2004interactionstrengths}.
However, these stability predictions are usually sensitive 
to the detailed assumptions of the mechanisms involved, the choice of functional response
(functional form of the relationship between interaction strength and population sizes),
and the choice of the large number of free parameters
\citep{williams2007yodzisinnes,gentleman2003zooplanktonATN,williams2004stability}.
Furthermore, due to the complexity of these models, studies that aim to derive general
features that characterize stable ecosystems have been focused on simulations
\citep{mccann1998weakinteractions,kokkoris1999weakinteractions,yodzis1981stability,ruiter1995stability}.
Their results therefore may not generalize well to systems beyond those that were simulated.

A parallel line of work for studying ecosystem stability assumes as little as possible about
detailed mechanisms and models species interactions with independent random variables
\citep{may1972stability,allesina2012stability}.
The simplicity of the model allows one to easily derive analytic predictions with great generality,
but the almost complete lack of mechanisms in the models means that they
are far from being realistic models of ecosystems. This could be a reason that these models
tend to give rise to counterintuitive predictions, e.g., larger systems are less stable
\citep{may1972stability} and realistic network structure is destabilizing in both predator-prey
and mutualitic bipartite networks \citep{allesina2012stability}.
Systems with many weak interactions are predicted to be less stable \citep{allesina2012stability},
even though simulations show otherwise
\citep{mccann1998weakinteractions,kokkoris1999weakinteractions}
and empirical works have shown that weak interactions pervade natural communities \citep{berlow2004interactionstrengths}.

In our work, we take an approach intermediate between the extreme of modeling ecological mechanisms
with as much detail as possible and the extreme of assuming almost nothing about the mechanisms involved,
aiming to achieve the generality obtainable from the latter while preserving at least some of the realism
achieved by the former.

Since energy flows are a crucial component of modeling population dynamics, we center our model of population
dynamics around a simple model of energy flow.
We focus on a community where interactions are dominated by feeding relationships between a \textit{consumer} trophic layer
and a \textit{resource} trophic layer. Given the allowed feeding interactions (the \textit{topology} of the food web)
and the aggregate energy flux through each species node, we derive the most probable energy flow rates along the
links of the food web as the solution to a simple set of equations. The equations are solved analytically
in the case of a perfectly nested topology.

The derivation relies on the principle of the maximization of information entropy (``MaxEnt''),
which allows one to make the least biased predictions consistent with some given prior information.
MaxEnt has led to successful theories in ecology
for patterns such as species-abundance and body-size distributions, species-area relationships,
bipartite food web degree distributions in quasistatic ecosystems,
and topological nestedness in mutualistic bipartite networks
\citep{harte2008ASNE,harte2015AGNSE,williams2010maxentdegree,payratoborras2019nestedness}.
Although our MaxEnt model of energy flows is not as accurate---model predictions are on average 0.3 orders of magnitude
apart from empirically observed flow rates---it still provides a reasonable approximation
given that the model is completely amechanistic.

The energy flows calculated from the MaxEnt model form the basis of the ecosystem's \textit{community matrix},
whose entries describe how changing the total biomass of one species affects the rate of change of the total biomass of another.
Deviations from the MaxEnt energy flows and processes not captured by energy flows (e.g., trophic efficiency)
are modeled as random variables. We then analytically investigate the stability characteristics of the resultant community matrix,
leveraging results from random matrix theory and perturbation theory. We discover mathematically that
\textit{nestedness promotes stability} for a quantitative metric of nestedness that we propose,
which we call the \textit{matrix dipole moment}.
The mathematical derivation is not a formal proof but a heuristic argument, so we also conduct
simulations to support the statement. We observe a correlation between 0.5 and 0.9
when $3 \leq S_R, S_C \leq 20$ and $(S_R, S_C) \neq (3, 3)$, where $S_R$ is the number of resource species
in the two-layer food web and $S_C$ is the number of consumer species.

\section*{Results}

\subsection*{\flow}

We model the distribution of flow rates from a resource (R) trophic layer to a consumer (C) trophic layer.
Here, the resource and consumer can be, for example, host and parasite, plant and herbivore, or herbivore and carnivore.
We construct from the flow rate distribution a corresponding joint probability distribution $f_{ij}$
proportional to the energy flow between resource species $i$ and consumer species $j$.
The total Shannon information entropy \citep{shannon1948entropy} of the distribution is then 
maximized under constraints derived from prior information.
The maximum-entropy distribution then corresponds to a least-biased,
most-probable energy flow rate distribution across the two trophic layers of interest.



Consider a normalized energy flow matrix $F$ with entries $f_{ij}$ denoting the energy flow from resource species $i$ to consumer species $j$, where $1 \leq i \leq S_R$ and $1 \leq j \leq S_C$. The normalization is such that the aggregate energy flux from resources to consumers is unity:
\begin{equation} \label{eq:norm}
    \sum_{i=1}^{S_R} \sum_{j=1}^{S_C} f_{ij} = 1.
\end{equation}
Let $E$ denote the set of permissible linkages $(i, j)$, where $1 \leq i \leq S_R$ and $1 \leq j \leq S_C$. In other words, $f_{ij} = 0$ if $(i, j) \not\in E$.
The first 6 entries of \cref{tab:variables} summarize the main variables defined in this subsection.

To apply MaxEnt, we need to specify a probability distribution whose information entropy is maximized under given constraints. The simplest way to construct a probability distribution from $f_{ij}$ is to simply reinterpret $f_{ij}$ itself as a probability distribution \citep{tanyimboh1993maxentflows}. To understand the physical meaning of the probability distribution $f_{ij}$, imagine energy flow as the flow of packets of unit energy. For any given time interval, if we randomly and uniformly choose one such energy packet from all those that flowed into the consumer trophic layer, then the probability that this packet flowed from resource species $i$ to consumer species $j$ is equal to $f_{ij}$.

Now, since $f_{ij}$ is a probability distribution, we can define its Shannon information entropy \citep{shannon1948entropy}:
\begin{equation} \label{eq:entropy}
    \r{H}(F^*) = -\sum_{(i,j)\in E} f_{ij} \ln f_{ij},
\end{equation}
where $F^*$ is the vector containing the normalized energy flows along permitted linkages.

To find the optimal energy flow distribution $F^*$, we maximize the entropy expression \cref{eq:entropy} under constraints imposed by prior information on aggregate energy flows ($d_i$ and $e_j$) through each species. Here, $e_j$ is the aggregate energy flow entering consumer node $j$ as a fraction of the total energy entering the consumer trophic layer ($\sum_j e_j = 1$). Similarly, $d_i$ is the aggregate energy flow exiting resource node $i$ as a fraction of the total energy exiting the resource trophic layer ($\sum_i d_i = 1$). Then, we have
\begin{subequations} \label{eq:constraints}
\begin{linenomath}
\begin{align}
    \sum_{\substack{j \\ (i,j) \in E}} f_{ij} &= d_i \label{eq:constraint1} \\
    \sum_{\substack{i \\ (i,j) \in E}} f_{ij} &= e_j. \label{eq:constraint2}
\end{align}
\end{linenomath}
\end{subequations}
Model flow rates are then obtained by maximizing the entropy expression \cref{eq:entropy} subject to the constraints \cref{eq:constraints}.

Solving the optimization problem yields
\begin{subequations} \label{eq:flowsol}
\begin{linenomath}
\begin{equation}
    \hat f_{ij} = a_i b_j, \label{eq:flowsol1}
\end{equation}
\end{linenomath}
where $a_i$ and $b_j$ satisfy
\begin{linenomath}
\begin{align}
    a_i \sum_{\substack{j \\ (i,j)\in E}} b_j &= d_i \label{eq:flowsol2} \\
    b_j \sum_{\substack{i \\ (i,j)\in E}} a_i &= e_j. \label{eq:flowsol3}
\end{align}
\end{linenomath}
\end{subequations}
Appendix S1 derives this result.

Note that \cref{eq:flowsol2} and \cref{eq:flowsol3} above list $S_R + S_C$ equations.
However, they are not independent: multiplying all $a_i$ by $c > 0$ and dividing all $b_j$ by $c$
leave \cref{eq:flowsol2,eq:flowsol3} unchanged.
Nevertheless, they generally give a unique positive solution in $f_{ij} = a_i b_j$
(or, there may be no positive solution; see below).
We can therefore fix one of the variables (e.g., $a_1 = 1$) and
numerically solve for the other $S_R + S_C - 1$ variables using $S_R + S_C - 1$ of the \cref{eq:flowsol2,eq:flowsol3}.


\subsubsection*{Perfectly nested topology}

Since nested bipartite subnetworks are common in food webs \citep{kondoh2010nestedsubwebs,cantor2017nestedness},
let us consider the case where our two-layer network is perfectly nested.

The adjacency matrix $A$ of the network is defined as $A_{ij} = \boldsymbol{1}[(i, j) \in E]$.
In other words, $A_{ij}$ is 1 if resource $i$ and consumer $j$ are connected and 0 otherwise.
Perfect nestedness means that there is an ordering of resource and consumer species such that
the 1s in the adjacency matrix occupy a contiguous region that includes the leftmost column and topmost row.

\begin{figure}[h!]
\centering
\includegraphics[width=0.75\textwidth]{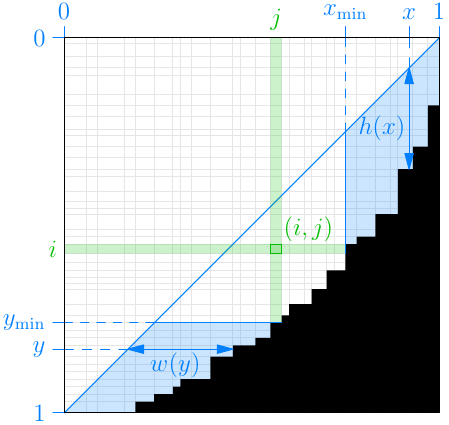}
\caption{Illustration of definitions of $y_\text{min}, x_\text{min}, w(y)$ and $h(x)$ in the MaxEnt solution to energy flows in the case of perfect topological nestedness.}
\label{fig:flowsoldefs}
\end{figure}

The MaxEnt energy flow solution in the perfectly nested case can be presented intuitively if we consider fitting the adjacency matrix
into a unit square (\cref{fig:flowsoldefs}), with the width of row $i$ equal to $d_i$ and the width of column $j$ equal to $e_j$.
The region where links are disallowed (0s in the adjacency matrix) is colored black.
Then the \textit{linkage existence boundary} (the line separating the white and black regions) will lie below the antidiagonal
(blue in \cref{fig:flowsoldefs}); Appendix S2 proves this statement.

To compute the MaxEnt energy flow from resource $i$ to consumer $j$,
let $x_\text{min}$ be the $x$-coordinate of where row $i$ meets the black region,
and let $y_\text{min}$ be the $y$-coordinate of where column $j$ meets the black region.\footnote{
$x$ increases from 0 to 1 from the left edge of the square to the right edge.
$y$ increases from 0 to 1 from the \textit{top} edge of the square to the \textit{bottom} edge.
}
For the strip between the anti-diagonal and the black region,
define $w(y)$ to be its horizontal width at distance $y$ from the top of the square,
and let $h(x)$ be its vertical height at distance $x$ from the left edge of the square.
Then the MaxEnt flow rates are given by the formula
\begin{equation} \label{eq:flowsolnested}
    \hat f_{ij} = C d_i e_j \exp\left(\int_{y_\text{min}}^{1} \frac{\text{d}y}{w(y)} + \int_{x_\text{min}}^{1} \frac{\text{d}x}{h(x)}\right),
\end{equation}
where $C$ is a normalization constant such that $\sum_{i,j} \hat f_{ij} = 1$.
Appendix S3 derives this result.

Suppose the strip has roughly uniform horizontal or vertical width, so that
we can write $w(y) \approx s$ and $h(x) \approx s$ for some $s$.
In addition, assume $d_i$ and $e_j$ satisfy energy equivalence \citep{brown2004energyequivalence}, i.e.,
$d_i \approx 1/S_R$ and $e_j \approx 1/S_C$.
Then $x_\text{min} \approx 1 + s - \frac{i}{S_R}$ and $y_\text{min} \approx 1 + s - \frac{j}{S_C}$,
so the MaxEnt flow rates $\hat f_{ij}$ increase exponentially
with increasing $\frac{i}{S_R} + \frac{j}{S_C}$ with rate $1/s$. In other words,
the flow rates increase exponentially towards the bottom-right and the increase is faster
when $s$ is smaller, i.e., the black region is closer to the anti-diagonal.

\subsection*{\empirical}

Although this paper focuses on the theoretical aspects of energy flow and stability,
it is worth testing the energy flow model against empirical data to make sure its predictions
are reasonable. Therefore, we tested our model against 14 empirical bipartite
food web subnetworks from 9 communities (table S1).

For each network studied, we calculated $d_i = \sum_j f_{ij}$ and $e_j = \sum_i f_{ij}$, where $f_{ij}$ are the normalized empirical flow rates. Then, we substituted these values for $d_i$ and $e_j$ into \cref{eq:flowsol} to obtain model flow rates $\hat f_{ij}$.
We have excluded species of degree 1 in our analysis as the flow through the linkage incident to such a species is always predicted with complete accuracy ($\hat f_{ij} = d_i = f_{ij}$ if the species is a resource and $\hat f_{ij} = e_i = f_{ij}$ if the species is a consumer).

\begin{figure}[h!]
\centering
\includegraphics[width=0.75\textwidth]{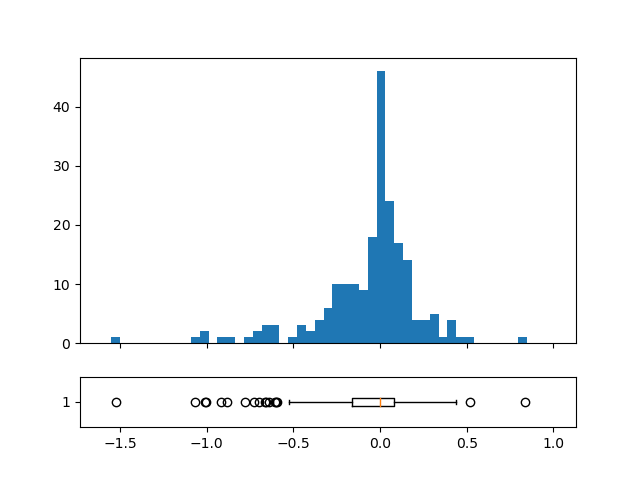}
\caption{Histogram and box-and-whisker plot of $\log_{10}(f_{ij}/\hat f_{ij})$. $\r{RMS = 0.294, Q1 = -0.164, Q3 = 0.078}$.}
\label{fig:flowratios}
\end{figure}

To visualize how well our model fits the data, we constructed a histogram and box plot for the distribution of $\log_{10}(f_{ij}/\hat f_{ij})$ (\cref{fig:flowratios}). 192 (91.4\%) out of the 210 values of $\log_{10}(f_{ij}/\hat f_{ij})$ lie between $-$0.5 and 0.5, the root mean square is 0.294, and the interquartile range (IQR) is 0.242. Our model thus predicts the order of magnitude of empirical observations accurately.

The box plot in \cref{fig:flowratios} shows 17 outliers (data points outside the interval $\r{[Q1 - 1.5IQR, Q3 + 1.5IQR]}$). 
This is expected as energy flow may depend on mechanisms specific to certain interspecific interactions
in natural communities, yet these mechanisms are neglected in the MaxEnt energy flow model.


\subsection*{\stability}

Based on the MaxEnt energy flow model proposed above, we model the dynamics of
a community where interactions are dominated by consumer-resource relationships
between two trophic layers.
Combining mathematical arguments and simulation,
we will show that a nested network is more likely to be stable.

One part of the argument assumes that the $d_i$ are similar in magnitude
and the same assumption is made for the $e_j$.
This roughly corresponds to the notion of energy equivalence in metabolic theory,
which has been shown both theoretically \citep{brown2004energyequivalence,harte2008ASNE,damuth2007energyequiv}
and empirically \citep{brown2004energyequivalence,damuth1981energyequiv,damuth1987energyequiv,marquet1990scaling}.

The main variables defined in this section are summarized in the last 4 entries of
\cref{tab:variables}.

\subsubsection*{A Nestedness Metric: Matrix Dipole Moment}

Since the notion of nestedness is informal, we first introduce a simple quantitative metric of nestedness which we call
the \textit{matrix dipole moment} ($N_p$).

Suppose the species are labeled such that the degree decreases with increasing $i$ and $j$.
Let's fit the adjacency matrix into a unit square like in \cref{fig:flowsoldefs},
such that row $i$ has width $d_i$ and column $j$ has width $e_j$.
The areas corresponding to a 1 in the adjacency matrix are given a uniform positive charge density
while the areas corresponding to a 0 are given a uniform negative charge density of the same magnitude.
Then $N_p$ is defined as the dipole moment in the $\nwarrow$ direction, normalized such that
the maximum possible value of $N_p$ is 1.

Note that perfect nestedness does not necessarily correspond to maximal $N_p$.
The $N_p$ value of a perfectly nested network is higher when
its linkage existence boundary is closer to the antidiagonal,
achieving the maximum of 1 when it follows the antidiagonal exactly.

While the concept of nestedness has traditionally been a property of the topology of the network only,
the definition presented above also depends on the aggregate energy flows $d_i$ and $e_j$.
If we assume energy equivalence such that $d_i = 1/S_R$ and $e_j = 1/S_C$, then we can write
the following formula for $N_p$ that is a function of the topology $E$ only:
\begin{equation} \label{eq:dipole}
    N_p = \frac{3}{S_R S_C} \sum_{i=1}^{S_R} \sum_{j=1}^{S_C} s_{ij} \left(1 - \frac{i-\frac12}{S_R} - \frac{j-\frac12}{S_C}\right),
\end{equation}
where
\[
    s_{ij} = \begin{cases}
        +1 & \text{if } (i, j) \in E \\
        -1 & \text{if } (i, j) \not\in E
    \end{cases}.
\]


\subsubsection*{Modeling Population Dynamics with Energy Flows}

For a community of $N$ species labeled $1, \ldots, N$, let $x_i$ be the aggregate
biomass of all individuals of species $i$. Then the dynamics of $\vec x = (x_1, \ldots, x_N)$
is modeled as a system of first-order differential equations of the form
\begin{equation} \label{eq:diff-eq}
    \frac{\r{d}\vec x}{\r{d}t} = f(\vec x).
\end{equation}
For full generality, we will not assume any particular form of the function $f(\vec x)$.
There is a rich literature on developing models for $f(\vec x)$ that incorporate various ecological mechanisms;
see Williams et al.\ (2007) \citep{williams2007yodzisinnes} for a review.

The community matrix $M_{ij}$ is defined as the Jacobian of $f(\vec x)$ at equilibrium, i.e.,
\begin{equation} \label{eq:community-def}
    M_{ij} = \left.\frac{\partial \dot x_i}{\partial x_j}\right|_{\vec x_0}
\end{equation}
where $f(\vec x_0) = 0$. Near $\vec x = \vec x_0$, the differential equations governing
the dynamics can be approximated as linear:
\begin{equation}
    \frac{\r{d}\vec x}{\r{d}t} \approx M(\vec x - \vec x_0).
\end{equation}
If all eigenvalues of $M$ have negative real parts, then starting close enough to $x_0$
will cause the system to converge to the equilibrium $\vec x_0$,
i.e., the equilibrium is stable. On the other hand, if $M$ has
an eigenvalue with a positive real part, then starting in the direction of the corresponding
eigenvector will cause the system to diverge away from the equilibrium, i.e.,
the equilibrium is unstable. Therefore, to study the stability of the system,
it suffices to study the eigenvalues of the community matrix $M$.

Without assuming any particular form of $f(\vec x)$, we will model $M$ as a random matrix,
following prior work \citep{may1972stability,allesina2012stability}.
However, whereas these works assume little in addition to the type of interaction studied,
we will incorporate energy flows as derived from the MaxEnt energy flow model.
In this case, the coefficients of the community matrix are split into a sum of 3 components:
\begin{equation} \label{eq:community-decomp}
    M = -dI + \tilde M + X.
\end{equation}
Self-interaction terms are modeled as a diagonal matrix $-dI$ ($d > 0$).
These terms represent increased competition for resources within a species as its population grows.
Consumer-resource interaction coefficients are given by the matrix $\tilde M$,
which we will call the \textit{consumer-resource community matrix}.
Letting the indices $1, \ldots, S_R$ denote the resources and the remaining
indices denote the consumers, we assume that the consumer-resource community matrix takes the form
\begin{equation} \label{eq:community-flows}
    \tilde M = \begin{bmatrix}
        0 & -\alpha F \\
        \beta(\eta \odot F)^\top & 0
    \end{bmatrix},
\end{equation}
where
\begin{itemize}
    \item $F$ is the matrix of energy flows $f_{ij}$. Energy flow rates $f_{ij}$ are
    MaxEnt flow rates \cref{eq:flowsol} multiplied by i.i.d.\ random variables with positive support.
    \item $\eta$ is the matrix of trophic efficiencies, defined such that
    only a fraction $\eta_{ij}$ of the energy flow $f_{ij}$ from resource $i$
    to consumer $j$ contributes to biomass growth of consumer $j$.
    The efficiencies $\eta_{ij}$ are modeled as random variables with positive support as well. 
    \item $\odot$ denotes the entry-wise matrix product (``Hadamard product'').
    \item $\alpha, \beta$ are positive constants.
\end{itemize}
The effect of changing consumer populations on resource populations
is described by entries of $-\alpha F$ and the effect of changing resource populations on consumer
populations is described by entries of $\beta(\eta \odot F)^\top$. 

The third term $X$ in \cref{eq:community-decomp} includes everything that is not modeled by the other two
terms. For example, it can include competition between different consumer species, mutualistic interactions,
interactions with other species in the ecosystem that are not modeled, etc.
Without any assumption on which of these mechanisms are present in the system
and how they are modeled, we will simply model $X$ as a random matrix whose entries are i.i.d.\ normal
random variables with mean 0 and variance $\sigma_X^2$. Since we are assuming that interactions in the system are dominated
by consumer-resource interactions described by $\tilde M$, $\sigma_X$ will be small compared to the entries
of $\tilde M$.

\subsubsection*{An Instability Metric for the Consumer-Resource Community Matrix}

The eigenvalues of $M$ are $\lambda_i = \lambda_i' - d$ where $\lambda_i'$ are the eigenvalues of $\tilde M + X$.
The largest real part $\max_i \re(\lambda_i')$ is the minimum value of $d$ such that the system is stable.
It therefore describes the degree to which the system is prone to being unstable.
Furthermore, by treating $X$ as a perturbation to $\tilde M$, we can obtain $\lambda_i'$ by adding first-order corrections
to eigenvalues $\tilde\lambda_i$ of $\tilde M$. These corrections have mean 0 and,
when the entries of $\eta$ are identical, their distribution is independent of $F$ for almost all $F$'s (Appendix S4.1.2).
This justifies the use of the largest real part among eigenvalues of $\tilde M$,
\begin{equation}
    m_{\tilde M} := \max_i \re(\tilde \lambda_i),
\end{equation}
as a measure of the \textit{in}stability of a consumer-resource community matrix $\tilde M$.

\subsubsection*{Nestedness Promotes Stability: The Mathematical Argument}

Here, we present a heuristic mathematical argument
that instability $m_{\tilde M}$ is typically smaller when the energy flow network $F$ underlying $\tilde M$
has higher nestedness $N_p$.

Without loss of generality, suppose $S_R \geq S_C$---the argument is similar when $S_R < S_C$.
Then the eigenvalues $\tilde\lambda_i$ of $\tilde M$ can be shown to contain $S_R - S_C$ zeros,
and the remaining $2S_C$ eigenvalues are $\pm \sqrt{-\alpha\beta\phi_j}$ where $\phi_j$ are the $S_C$ eigenvalues
of $(\eta \odot F)^\top F$ (Appendix S4.1.1). If all $\phi_j$ are non-negative, then all $\tilde\lambda_i$ are purely imaginary
and $m_{\tilde M} = 0$.
When all entries of $\eta$ are the same positive value $\overline\eta$,
then $(\eta \odot F)^\top F = \overline\eta F^\top F$ is positive semi-definite,
so all $\phi_i$ are non-negative.
Therefore, a positive real part in some $\tilde\lambda_i$ only arises due to fluctuations in the entries of $\eta$,
resulting in negative or non-real values of $\phi_j$. This then causes some
$\tilde\lambda_i = \pm\sqrt{-\alpha\beta\phi_j}$ to be positive, as a result of which the instability $m_{\tilde M}$ becomes positive.

Therefore, stability is enhanced when the positive semi-definiteness of $\overline\eta F^\top F$ is more robust
to perturbations due to fluctuations in the entries of $\eta$.
In other words, \textit{stability is more likely when the smallest eigenvalue of $F^\top F$ is larger}.
This heuristic measure for the stability of an energy flow matrix $F$ constitutes the core of our argument.

We will consider two separate cases: the case where the topology is perfectly nested and the case where it is not.

\paragraph*{Perfectly nested topology}
The argument below assumes approximate energy equivalence, i.e., the $d_i$ are not too different from each other
and neither are the $e_j$.

Let's first consider the case where
all consumers consume different numbers of resources. Then the MaxEnt energy flow solution $F$ is a non-singular matrix.
According to the discussion that follows \cref{eq:flowsolnested},
the entries of $F$ decrease exponentially towards the top-left corner, and this exponential
decay is faster when the linkage existence boundary is closer to the antidiagonal, i.e., $N_p$ is higher.
Thus, $\left(F^\top F\right)_{ij} = \vec f_i \cdot \vec f_j$ ($\vec f_k$ denotes column $k$ of $F$) decreases exponentially
in $|i - j|$, with the exponential decay faster with higher values of $N_p$ as well.
In other words, when the topology is more nested according to $N_p$,
the entries of $F^\top F$ are more concentrated on the diagonal,
thus resulting in larger eigenvalues that are less spread out (Appendix S4.2).
This causes the smallest eigenvalue
of $F^\top F$ to be larger, resulting in more stability.

When there are consumers that consume the same number of resources, then their corresponding columns
in the MaxEnt flow rate matrix are equal and the smallest eigenvalue of $F^\top F$ is zero.
However, this is rarely the case in real food webs,
where entries of $F$ typically deviate from the MaxEnt solution by about 0.3 orders of magnitude (see subsection \empirical).
When these deviations are modeled as independent random variables that are added to the MaxEnt solution
as a perturbation, it can be shown that the original zero eigenvalues of $F^\top F$ become positive,
with the smallest one typically larger for larger $N_p$ (Appendix S4.3).
 
\begin{figure}[h!]
\centering
\includegraphics[width=0.75\textwidth]{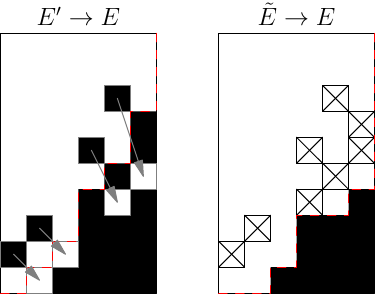}
\caption{Left: Non-nested topology $E$ is generated by moving 1s in the adjacency matrix (white cells)
of nested topology $E'$ into the black region.
Right: $E$ is generated by changing 1s in the adjacency matrix of nested topology $\tilde E$
to 0s, shown as crossing out white cells.
The red dashed lines are the linkage existence boundaries of $E'$ and $\tilde E$, respectively.}
\label{fig:nonnested}
\end{figure}

\paragraph*{Non-perfectly-nested topology} Here, we provide a heuristic argument that a non-perfectly-nested
topology tends to be more stable for larger $N_p$.
We compare two networks that share the same number of edges, $d_i$ and $e_j$, but with different topologies:
$E$, a non-perfectly-nested topology, and $E'$, a perfectly nested one.
Suppose that $E$ can be obtained from $E'$ by moving 1s in the adjacency matrix of $E'$
towards the bottom-right corner (left panel of \cref{fig:nonnested}), thus decreasing $N_p$.
Our goal now is to argue that a community with food web topology $E$ is less likely to be stable than
one with topology $E'$.

Suppose that $E$ is adequately represented as a typical sample from the distribution
of random topologies generated as follows:
from a perfectly nested topology $\tilde E$, keep the $(i, j)$-edge with probability $p_i q_j$
where $0 < p_i, q_j < 1$ (right panel of \cref{fig:nonnested}).
The linkage existence region of $\tilde E$ should contain that of $E$, so that its linkage existence
boundary is towards the bottom-right of that of $E'$ (\cref{fig:nonnested}).
Suppose $\tilde f_{ij} = \tilde a_i \tilde b_j$ are the MaxEnt flow rates of $\tilde E$ with the same $d_i, e_j$ as $E$ and $E'$
(see \cref{eq:flowsol}).
Then $a_i = \tilde a_i/p_i, b_j = \tilde b_j/q_j$ solve \cref{eq:flowsol2,eq:flowsol3} in expectation,
in the sense that the expectation values of the left-hand sides over the randomness in $E$
equal the corresponding right-hand sides.
As a result, $f_{ij} = a_i b_j = \tilde f_{ij}/p_i q_j$ approximate the MaxEnt flow rates of the original network.
Using this approximation, we obtain
\[
\mathbb E[(F^\top F)_{ij}] = \mathbb E\left[\sum_k \boldsymbol{1}[(i,k) \in E)]\boldsymbol{1}[(j,k) \in E)] \frac{\tilde f_{ik}}{p_i q_k} \frac{\tilde f_{jk}}{p_j q_k}\right] = \sum_k \tilde f_{ik} \tilde f_{jk} = (\tilde F^\top \tilde F)_{ij}.
\]
Thus, the eigenvalues of $F^\top F$ are expected to be distributed similarly to those of $\tilde F^\top \tilde F$,
so the stability of $E$ should be similar to that of $\tilde E$. Yet $\tilde E$ is less stable than $E'$
since both are perfectly nested and the latter has a linkage existence boundary closer to the anti-diagonal.
Thus, $E$ is less stable than $E'$, as desired.


\subsubsection*{Nestedness Promotes Stability: The Simulation}

For each $(S_R, S_C)$ with $3 \leq S_R, S_C \leq 20$, we generated $10 S_R S_C$ random topologies.
The sampling algorithm is described in Appendix S5 and is designed to generate a wide diversity of nestedness values. 
For each random topology, the MaxEnt solution is computed according to \cref{eq:flowsol} with $d_i = 1/S_R$ and $e_j = 1/S_C$.
Randomness is introduced by multiplying each flow rate by an independent random variable
following the distribution of $10^X$ where $X$ is a Gaussian
with standard deviation matching the empirically estimated value of 0.3 (see subsection \empirical).
The entries of the efficiency matrix $\eta$ were chosen
to be exponentials of i.i.d.\ Gaussian random variables with standard deviation $0.5$
to approximate a realistic distribution of efficiencies. Since scaling the entire efficiency
matrix simply scales all eigenvalues of $(\eta \odot F)^\top F$, we chose the Gaussian
random variables to have mean 0 without loss of generality.

\begin{figure}[h!]
\centering
\includegraphics[width=0.75\textwidth]{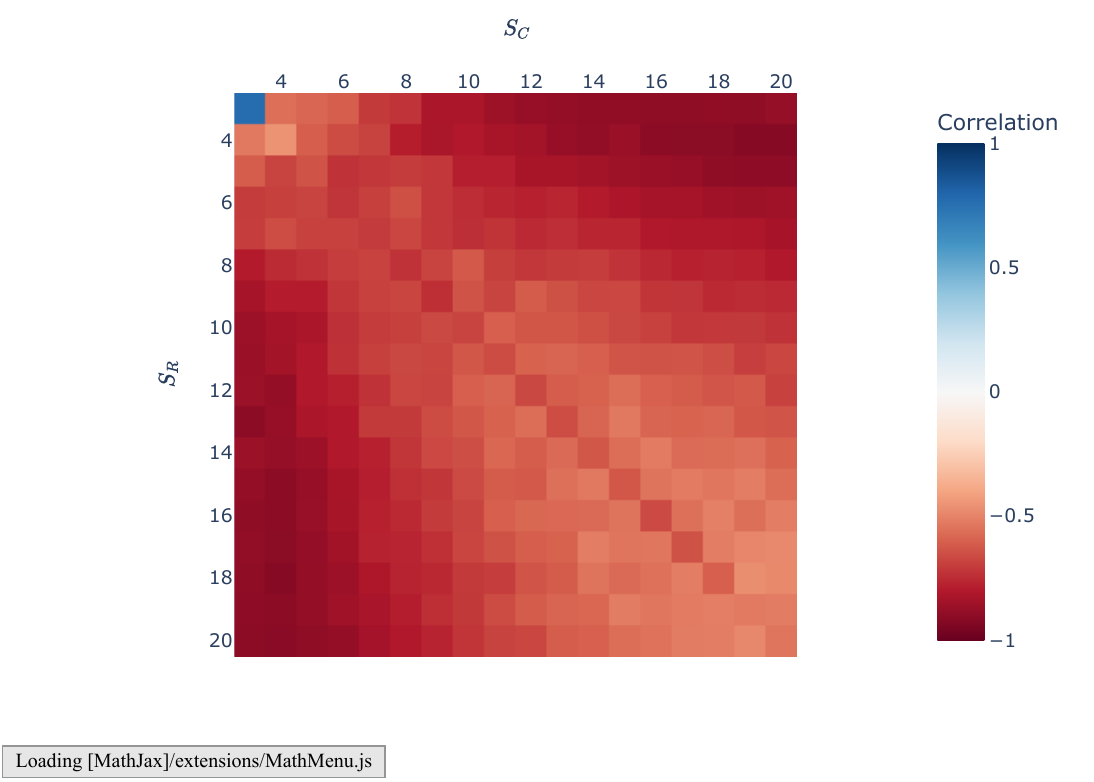}
\caption{Heat map showing the correlation between instability ($\overline m$) and nestedness ($N_p$)
for each $(S_R, S_C)$ pair with $3 \leq S_R, S_C \leq 20$.}
\label{fig:stability}
\end{figure}

For each topology, we generated 1000 pairs $(F, \eta)$ as described above
and computed $m = \max_j \re(\sqrt{-\phi_j})$ where $\phi_j$ are the eigenvalues of $(\eta \odot F)^\top F$,
so that $\sqrt{\alpha\beta}m$ is the maximum real part of $\tilde M$,
the part of the community matrix due to consumer-resource interactions.
We use the mean $\overline m$ of all the $m$ values as our metric of \textit{in}stability of the given topology.
A linear regression is conducted between $\overline m$ and $N_p$ (\cref{eq:dipole}), and the resultant
correlation values for all $(S_R, S_C)$ pairs with $3 \leq S_R, S_C \leq 20$ are plotted on a heat map (\cref{fig:stability}).
All linear regressions give nonzero slopes (all $p$-values are less than \SI{e-9}{}),
and correlations range between $-0.5$ and $-0.9$ with the sole exception of $(S_R, S_C) = (3, 3)$.
Correlation is stronger when $S_R$ and $S_C$ are more different: it is less than $-0.7$ when
$S_R \geq 2S_C$ or $S_C \geq 2S_R$.

%


To visualize the relationship between the instability metric $\overline m$
and the nestedness metric $N_p$, we plotted $\overline m$
against $N_p$ for 600 topologies generated with $S_R = 10, S_C = 6$ (\cref{fig:stability_ex}).
Despite some scatter, the negative trend is manifest.

In Appendix S6, we investigate the relationship between instability and nestedness
using 4 nestedness metrics ($N_p$ plus three other metrics from the literature) and 2 instability metrics
($\overline m$ and an additional metric $p_{m > 0}$, defined as the proportion of $m$
values that are positive).
We consistently obtain negative correlation betwen instability and nestedness when $S_R$ and $S_C$
are sufficiently large and not too close to each other.
The correlation between stability and nestedness is not as strong with the three existing nestedness metrics
as it is with $N_p$, which is expected
since the argument that nestedness promotes stability is specific to the $N_p$ nestedness measure.

\section*{\discussion}

Traditionally, theoretical approaches to stability often assumed very little about the system
\citep{may1972stability,allesina2012stability}
or attempted to model the system in as much detail as possible
\citep{williams2004stability}. Our approach is intermediate between the two.
While it attempts to somewhat realistically model real systems by incorporating
some mechanisms (in our case, the effect of energy flow on community matrix coefficients),
the model remains simple enough that analytic mathematical arguments can be applied, even for large systems.
As a result, compared to approaches that model the system in detail, which are often restricted to small systems
of only a few species, we were able to break this restriction and make predictions about systems of arbitrary size.
Our model also has a lot fewer free parameters than models based on functional response.
In fact, the only adjustable parameters in the simulation are the algorithm for generating random topologies
and the distributions of the random variables
used to construct the community matrix (efficiency values and perturbations to the MaxEnt energy flow rates).
Therefore, whereas stability predictions from functional response models
tend to be sensitive to the parameters and specific functional forms of the functional response
\citep{williams2007yodzisinnes,gentleman2003zooplanktonATN,williams2004stability},
predictions from our model are more robust.

Also, unlike the stability studies that assumed little about the system \citep{may1972stability,allesina2012stability},
our results are consistent with empirically established features that characterize ecosystems.
One general feature of ecological communities is ``many weak interactions''
\citep{berlow2004interactionstrengths,ruiter1995stability,neutel2002stability},
which have been shown to promote stability in simulations \citep{mccann1998weakinteractions,kokkoris1999weakinteractions}
as well as empirical food webs \citep{yodzis1981stability,ruiter1995stability}.
When $N_p$ is high, high values of energy flow are produced right below the antidiagonal of the energy flow matrix
and the region above the antidiagonal is filled with exponentially smaller energy flows, reproducing
the phenomenon of ``many weak interactions.'' Another tendency of empirical food webs is intervality,
the property that the species can be ordered such that each predator's prey are contiguous in the ordering
\citep{stouffer2006intervality}. We note that interval food webs arise from a high-$N_p$ MaxEnt flow network
when the weakest links---those in the top-left corner of the adjacency matrix---are dropped
(e.g., because they are below the measurement threshold).
Thus, our finding that networks with high $N_p$ tend to be more stable is consistent with
the prevalence of interval food webs observed in real ecosystems.
Finally, it is worth noting that the pattern where interactions are stronger near the linkage-existence
boundary of a nested network has been shown to promote stability in bipartite mutualistic networks \citep{staniczenko2013}.
Our results can therefore be viewed as an extension of this trend to two-layer food webs.

Our results provide a partial explanation for stability in communities
dominated by consumer-resource interactions between two trophic layers.
First, a nested topology is generated by known mechanisms, such as
body size constraints
\citep{ings2009nestedbodysize,woodward2007bodysize,yvondurocher2008,cohen1993bodysize,warren1987bodysize}
and phylogenetic constraints \citep{ings2009nestedbodysize,cattin2004phylogeneticnestedness}.
Then, according to the results derived in this paper, the energy flows typical for a nested topology
give rise to population dynamics that are more likely to have a stable equilibrium.

\section*{Conclusion}

We constructed a model of population dynamics in a community dominated
by trophic interactions between two trophic layers.
The main assumptions are the maximum entropy principle for modeling energy flows underlying the population dynamics,
and approximate energy equivalence, which can also be derived from MaxEnt \citep{harte2008ASNE}.
We introduced a new quantitative nestedness metric that is simple and intuitive---the matrix dipole moment $N_p$.
It is a good predictor for stability in our model of population dynamics,
as shown through both heuristic mathematical arguments and simulations.

Our results show that incorporating a small number of mechanisms---fewer than ``fully fleshed-out'' models that specify
the functional response of all interactions, but more than ``bare-bones'' models that model interactions as
i.i.d.\ random variables---can give rise to a minimally realistic model that
yields powerful and general analytic predictions. We encourage future researchers to
further explore this kind of approach in ecological modeling.

\newpage

\bibliographystyle{naturemag}
\bibliography{refs}

\newpage{}

\section*{Tables}
\renewcommand{\thetable}{\arabic{table}}
\setcounter{table}{0}

\begin{table}[h!]
\caption{List of important variables defined in this paper.}
\label{tab:variables}
\centering
\begin{spacing}{1.5}
\begin{tabu}{r|[2pt]p{5in}}
Variable & Definition \\ \tabucline[2pt]{-}
$S_R$ & Number of resource species in the two-layer food web \\ \hline
$S_C$ & Number of consumer species in the two-layer food web \\ \hline
$E$ & The topology of the food web as a set of pairs $(i, j)$ where resource $i$ ($1 \leq i \leq S_R$) is connected to consumer $j$ ($1 \leq j \leq S_C$) \\ \hline
$f_{ij}$ & Energy flow rate from resource $i$ to consumer $j$, normalized according to \cref{eq:norm} \\ \hline
$d_i$ & Aggregate energy flow rate exiting resource species $i$, normalized such that $\sum_i d_i = 1$ \\ \hline
$e_j$ & Aggregate energy flow rate entering consumer species $j$, normalized such that $\sum_j e_j = 1$ \\ \hline
$N_p$ & The matrix dipole moment, the new nestedness measure introduced in this paper; see \cref{eq:dipole} \\ \hline
$M$ & The $(S_R + S_C) \times (S_R + S_C)$ community matrix, defined as the Jacobian matrix at equilibrium of the first-order ODE system modeling population dynamics \\ \hline
$\tilde M$ & The contribution to the community matrix due to consumer-resource interactions in the two-layer food web; see \cref{eq:community-flows} \\ \hline
$m_{\tilde M}$ & The largest real part of any eigenvalue of $\tilde M$; measures the instability of the system
\end{tabu}
\end{spacing}
\end{table}

\newpage{}
\newpage{}

\section*{Figures}



\newpage


\newpage

\newpage

\newpage

\begin{figure}[h!]
\centering
\includegraphics[width=0.75\textwidth]{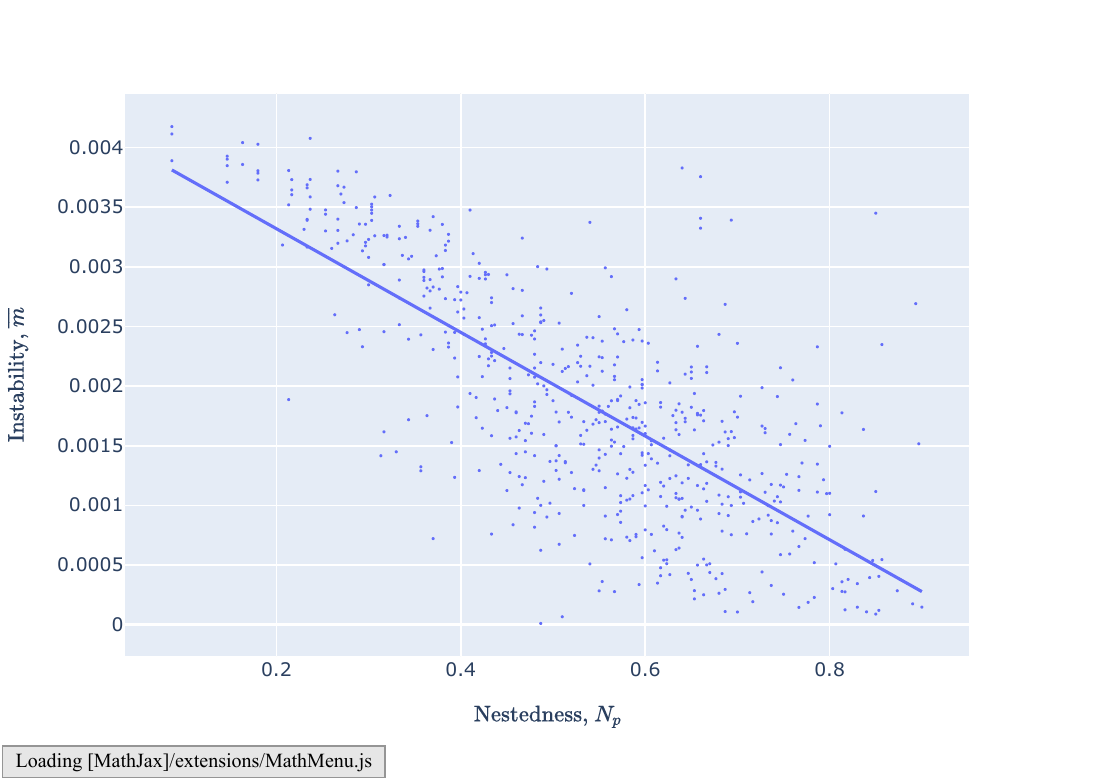}
\caption{Plot of instability metric $\overline m$ versus nestedness metric $N_p$
for 600 randomly sampled topologies with $S_R = 10, S_C = 6$.}
\label{fig:stability_ex}
\end{figure}
\newpage





%
%


\newpage

\section*{Acknowledgments}

Funding for J.H. was provided by grant DEB 1751380 from the US National Science Foundation.
We thank P. P. A. Staniczenko and N. D. Martinez for useful discussions.

\section*{Statement of Authorship}

J.H.\ and Z.L.\ provided ideas for the MaxEnt modeling of energy flows. Z.L.\ explored the models, derived the analytic solution and its stability implications, and conducted empirical testing. Z.L.\ wrote the paper. Z.L.\ and J.H.\ edited the paper.

\section*{Data and Code Availability}

This submission uses novel code, provided at the following 
GitHub repositories:
\url{https://github.com/uranium11010/nestedness-stability}
(stability simulations);
\url{https://github.com/uranium11010/network-flow-model}
(flow model empirical testing).

\section*{Additional Information}

Supplementary Information is available for this paper.


\newpage

\renewcommand{\theequation}{S\arabic{equation}}
\renewcommand{\thetable}{S\arabic{table}}
\renewcommand{\thesection}{S\arabic{section}}
\renewcommand{\thefigure}{S\arabic{figure}}
\setcounter{equation}{0}  
\setcounter{figure}{0}
\setcounter{table}{0}

\section{Derivation of the MaxEnt Energy Flow Solution for General Topology}

The Lagrangian is given by
\begin{equation}
    \mathcal{L}\left( F^{*},\boldsymbol{\kappa},\boldsymbol{\lambda} \right) = - \sum_{i,j}^{}{f_{ij}\ln f_{ij}} - \sum_{i}^{}{\kappa_{i}\left( \sum_{j}^{}f_{ij} - d_{i} \right)} - \sum_{j}^{}{\lambda_{j}\left( \sum_{i}^{}f_{ij} - e_{j} \right)}, \label{Seq:dis-lagrangian}
\end{equation}
where we didn't include the normalization condition eq.~(1) as a
constraint as it is already implied by the constraints eq.~(3a) and eq.~(3b).

We now need to solve for
\begin{equation}
\nabla\mathcal{L} = 0. \label{Seq:dis-stationary}
\end{equation}
Note that \(\frac{\partial\mathcal{L}}{\partial\kappa_{i}} = 0\) and
\(\frac{\partial\mathcal{L}}{\partial\lambda_{j}} = 0\) recover the
constraint equations eq.~(3a) and eq.~(3b). On the other
hand, \(\frac{\partial\mathcal{L}}{\partial f_{ij}} = 0\) gives
\begin{linenomath}
\begin{align}
    - \ln f_{ij} - 1 - \kappa_{i} - \lambda_{j} &= 0 \notag \\
    f_{ij} &= e^{- \kappa_{i} - \lambda_{j} - 1} \notag \\
    f_{ij} &= a_{i}b_{j} \label{Seq:dis-factorized}
\end{align}
\end{linenomath}
where \(a_{i} \equiv e^{- \kappa_{i} - \frac{1}{2}} > 0\) and
\(b_{j} \equiv e^{- \lambda_{j} - \frac{1}{2}} > 0\).

\section{Proof That the Linkage Existence Boundary Lies Below the Antidiagonal}

Suppose there's a point on the linkage existence boundary that is above the antidiagonal.
Let that point be the intersection of horizontal line $l_h$ and vertical line $l_v$.
Then the total flow below $l_h$ (equal to its distance to the bottom side of the square)
is larger than the total flow to the left of $l_v$ (equal to its distance to the left side of the square),
which is impossible since the former is a strict subset of the latter.

\section{Derivation of the MaxEnt Energy Flow Solution for a Perfectly Nested Topology}

Let's verify that the MaxEnt energy flow formula given in the main text is correct.

Since $x_\text{min}$ only depends on $i$ and $y_\text{min}$ only depends on $j$,
we can factorize $\hat f_{ij} = a_i b_j$ where
\[
a_i \propto d_i \exp\left(\int_{x_\text{min}}^1 \frac{\d x}{h(x)}\right) \quad \text{and} \quad
b_j \propto e_j \exp\left(\int_{y_\text{min}}^1 \frac{\d y}{w(y)}\right).
\]
It now suffices to verify
\[
    \sum_{i: (i, j) \in E} \hat f_{ij} = e_j \quad \text{and} \quad
    \sum_{j: (i, j) \in E} \hat f_{ij} = d_i.
\]
Due to the symmetry between the indices $i$ and $j$, we only need to show the former,
and the proof for the latter would be nearly identical.

Substituting the expression for $\hat f_{ij}$ gives the following statement that we have to verify:
\begin{equation} \label{Seq:constraintsum}
    \sum_{i: (i, j) \in E} d_i \exp\left(\int_{y_\text{min}}^{1} \frac{\d y}{w(y)} + \int_{x_\text{min}}^{1} \frac{\d x}{h(x)}\right) = 1/C
\end{equation}
is constant in $i$.

Let's define $x_m(y)$ to be the maximum $x$ such that $(x, y)$ is in the white region (including the boundary) of the square,
and $y_m(x)$ is similarly defined as the maximum $y$ such that $(x, y)$ is in the white region (including the boundary) of the square.
Furthermore, for any $x$ in column $j$ (excluding its left and right edges), $h(x) = y_m(x) + x - 1$;
and for any $y$ in row $i$ (excluding its top and bottom edges), $w(y) = x_m(y) + y - 1$.

For any $x$ in column $j$ (excluding its left and right edges), the left side of \cref{Seq:constraintsum} can be rewritten as
\[
    \int_0^{y_m(x)} \exp\left(\int_{y_m(x)}^{1} \frac{\d y'}{x_m(y') + y' - 1} + \int_{x_m(y)}^{1} \frac{\d x'}{y_m(x') + x' - 1}\right)\d y
\]
or
\begin{equation} \label{Seq:constraintint}
    \exp\left(\int_{y_m(x)}^{1} \frac{\d y}{x_m(y) + y - 1}\right) \int_0^{y_m(x)} \exp\left(\int_{x_m(y)}^{1} \frac{\d x'}{y_m(x') + x' - 1}\right)\d y,
\end{equation}
which we wish to show to be independent of $x$ over all of $[0, 1]$ except for those points at which $y_m(x)$ is discontinuous.

To avoid issues that may arise due to these discontinuities, let's write the linkage existence boundary
as a parameterized curve $(x(t), y(t))$ for $0 \leq t \leq 1$, which goes from $(1, 0)$ in the top-right corner
at $t = 0$ to $(0, 1)$ in the bottom-left corner at $t = 1$. The parameterization is such that the curve is
continuously differentiable everywhere except at the corners, where it is only continuous.

Consider applying integration by parts to the following integral:
\begin{linenomath}
\begin{align*}
    &\phantom{=\ } \int_0^{t_0} (\dot x(t) + \dot y(t)) \exp\left(\int_{t}^0 \frac{\dot x(t')}{x(t') + y(t') - 1}\,\d t'\right)\d t \\
    &= (x(t_0) + y(t_0) - 1) \exp\left(\int_{t_0}^0 \frac{\dot x(t')}{x(t') + y(t') - 1}\,\d t'\right) \\
    &\qquad - \int_0^{t_0} (x(t) + y(t) - 1) \exp\left(\int_{t}^0 \frac{\dot x(t')}{x(t') + y(t') - 1}\,\d t'\right) \frac{-\dot x(t)}{x(t) + y(t) - 1}\,\d t \\
    &= (x(t_0) + y(t_0) - 1) \exp\left(\int_{t_0}^0 \frac{\dot x(t)}{x(t) + y(t) - 1}\,\d t\right) + \int_0^{t_0} \dot x(t) \exp\left(\int_{t}^0 \frac{\dot x(t')}{x(t') + y(t') - 1}\,\d t'\right) \d t.
\end{align*}
\end{linenomath}
Moving the integral on the right-hand side to the left side cancels out the $\dot x(t)$ in the integrand on the left-hand side,
giving
\[
    \int_0^{t_0} \dot y(t) \exp\left(\int_{t}^0 \frac{\dot x(t')}{x(t') + y(t') - 1}\,\d t\right)\d t = (x(t_0) + y(t_0) - 1) \exp\left(\int_{t_0}^0 \frac{\dot x(t)}{x(t) + y(t) - 1}\,\d t\right).
\]
Therefore, we obtain
\begin{linenomath}
\begin{multline} \label{Seq:constraintintsol}
    S(t_0) := \exp\left(\int_{t_0}^1 \frac{\dot y(t)}{x(t) + y(t) - 1}\,\d t\right) \int_0^{t_0} \dot y(t) \exp\left(\int_{t}^0 \frac{\dot x(t')}{x(t') + y(t') - 1}\,\d t'\right)\d t \\
    = (x(t_0) + y(t_0) - 1) \exp\left(\int_{t_0}^0 \frac{\dot x(t)}{x(t) + y(t) - 1}\,\d t + \int_{t_0}^1 \frac{\dot y(t)}{x(t) + y(t) - 1}\,\d t\right).
\end{multline}
\end{linenomath}
Note that $S(t_0)$ becomes \cref{Seq:constraintint}
when we set $t_0$ to the number such that $x(t_0) = x$.

It now suffices to show that $S(t_0)/S(t_1) = 1$ for any $0 < t_0, t_1 < 1$. Indeed,
\begin{linenomath}
\begin{align*}
    \frac{S(t_0)}{S(t_1)} &= \frac{x(t_0) + y(t_0) - 1}{x(t_1) + y(t_1) - 1} \exp\left(\int_{t_0}^{t_1} \frac{\dot x(t)}{x(t) + y(t) - 1}\,\d t + \int_{t_0}^{t_1} \frac{\dot y(t)}{x(t) + y(t) - 1}\,\d t\right) \\
    &= \frac{x(t_0) + y(t_0) - 1}{x(t_1) + y(t_1) - 1} \exp\left(\int_{t_0}^{t_1} \frac{\dot x(t) + \dot y(t)}{x(t) + y(t) - 1}\,\d t\right) \\
    &= \frac{x(t_0) + y(t_0) - 1}{x(t_1) + y(t_1) - 1} \exp\left(\ln(x(t) + y(t) - 1)\big|_{t_0}^{t_1}\right) \\
    &= 1,
\end{align*}
\end{linenomath}
hence completing the proof.

\section{Details of the Stability Argument}

\subsection{The spectrum and perturbation theory of $\tilde M$}

\subsubsection{Spectrum of $\tilde M$} \label{Ssec:spectrum}

Let's write $\tilde M = \begin{bmatrix} 0 & A \\ B & 0 \end{bmatrix}$ where $A = -\alpha F$ and $B = (\eta \odot F)^\top$.

Let's first look at the relationship between the eigenvalues/eigenvectors of $AB$ and those of $BA$.
For any $\mu \in \mathbb C$, let $T_\mu$ be the $\mu$-eigenspace of $AB$,
where $T_\mu = \{0\}$ if $\mu$ is not an eigenvalue of $AB$. Similarly,
let $S_\mu$ be the $\mu$-eigenspace of $BA$. We notice the following:
\begin{itemize}
    \item If $\mu \neq 0$, then $ABT_\mu = T_\mu$ and $BAS_\mu = S_\mu$.
    Thus, $T_\mu$ and $S_\mu$ are related by $T_\mu = AS_\mu$ and $S_\mu = BT_\mu$,
    and $\dim(T_\mu) = \dim(S_\mu)$.
    As a result, $\mu$ is an eigenvalue of $AB$ iff it is an eigenvalue of $BA$.
    \item If $\mu = 0$, then $T_0$ is just the kernel of $AB$ and $S_0$ is the kernel of $BA$.
\end{itemize}

Now, $\tilde M^2 = \begin{bmatrix} AB & 0 \\ 0 & BA \end{bmatrix}$, whose $\mu$-eigenspace
is just $T_\mu \oplus S_\mu$.
\begin{itemize}
    \item For a nonzero eigenvalue $\mu$ of $\tilde M^2$,
    the possible corresponding eigenvalues of $\tilde M$ are $\lambda = \pm\sqrt{\mu}$
    and the corresponding eigenvectors take the form $v \oplus u$ where $v \in T_\mu$ and $u \in S_\mu$.
    Now, writing $\tilde M(v \oplus u) = \lambda(v \oplus u)$ gives $Au = \lambda v$ and $Bv = \lambda u$,
    so the eigenspace of $\tilde M$ corresponding to an eigenvalue $\lambda = \pm\sqrt{\mu} \neq 0$ is
    $\{(\lambda v) \oplus (Bv) : v \in T_{\mu}\}$.
    \item For an eigenvalue $\mu = 0$ of $\tilde M^2$ (if one exists),
    the corresponding eigenvalue of $\tilde M$ is $\lambda = \pm\sqrt{\mu} = 0$
    and the corresponding eigenvectors take the form $v \oplus u$ where $v$ is in the kernel of $B$
    and $u$ is in the kernel of $A$.
\end{itemize}

\subsubsection{Perturbation theory of $\tilde M$}

\textit{(Note: Every square matrix in this section is assumed to be diagonalizable.)}

Consider adding to $\tilde M$ a random perturbation $X$ whose entries are
i.i.d.\ normal with mean 0 and variance $\sigma^2$ that is small.
This perturbs the eigenvalues $\tilde\lambda_i$ of $\tilde M$ such that they become
$\lambda_i' = \tilde\lambda_i + \delta\tilde\lambda_i$. Let's derive a formula
for $\delta\tilde\lambda_i$ to first order in $X$ to show that the mean is 0
and, when the entries of $\eta$ are constant, the joint distribution is independent of $F$ when all
its singular values are distinct and positive.

For an eigenvalue $\tilde\lambda$ of $\tilde M$ with multiplicity $m \geq 1$,
let $\tilde v_1, \tilde v_2, \ldots, \tilde v_m$ be a basis for its eigenspace $D_{\tilde\lambda}$.
Since $\tilde M$ and $\tilde M^\top$ have the same characteristic polynomial,
$\tilde M$ also has a left eigenspace $E_{\tilde\lambda}^*$
spanned by $\tilde w_1^\dagger, \tilde w_2^\dagger, \ldots, \tilde w_m^\dagger$, corresponding to eigenvalue $\tilde\lambda$.

Under the perturbation $X$,
the perturbed eigenvectors can be decomposed as $\tilde v_i' + \delta\tilde v_i$, where $\tilde v_i' \in D_{\tilde\lambda}$
and $\delta\tilde v_i \perp E_{\tilde\lambda}$.\footnote{Here,
we have assumed that $D_{\tilde\lambda} \oplus E_{\tilde\lambda}^\perp = V$,
where $V = \mathbb C^{S_R + S_C}$ is the entire vector space.
This assumption should generally hold since $\dim D_{\tilde\lambda} = m$ and
$\dim E_{\tilde\lambda}^\perp = S_R + S_C - \dim E_{\tilde\lambda} = S_R + S_C - m$.
Also, we know that it holds for Hermitian matrices where $D_{\tilde\lambda} = E_{\tilde\lambda}$.}
The deviations $\delta\tilde v_i$ from the eigenspace can be considered to be small, so
\begin{linenomath}
\begin{align}
    (\tilde M + X)(\tilde v_i' + \delta\tilde v_i) &= (\tilde\lambda_i + \delta\tilde\lambda_i)(\tilde v_i' + \delta\tilde v_i) \notag \\
    \tilde M\,\delta\tilde v_i + X\tilde v_i' &= \tilde\lambda_i\,\delta\tilde v_i + (\delta\tilde\lambda_i)\tilde v_i', \label{Seq:pertdeg}
\end{align}
\end{linenomath}
where second-order terms have been neglected.
Multiplying both sides of \cref{Seq:pertdeg} by $\tilde w_j^\dagger$ from the left gives
\begin{linenomath}
\begin{align}
    \tilde\lambda\tilde w_j^\dagger\delta\tilde v_i + \tilde w_j^\dagger X\tilde v_i' &= \tilde\lambda\tilde w_j^\dagger\delta v_i + \tilde w_j^\dagger(\delta\tilde\lambda)\tilde v_i' \notag \\
    \tilde w_j^\dagger X\tilde v_i' &= \tilde w_j^\dagger\tilde v_i'(\delta\tilde\lambda). \label{Seq:perteigdeg}
\end{align}
\end{linenomath}
Let's define the matrices
\[
    W = \begin{bmatrix} w_1 & w_2 & \ldots & w_m \end{bmatrix} \qquad
    V = \begin{bmatrix} v_1 & v_2 & \ldots & v_m \end{bmatrix} \qquad
    VV' = \begin{bmatrix} v_1' & v_2' & \ldots & v_m' \end{bmatrix},
\]
where column $i$ of $V'$ is just the coefficients of writing $v_i'$ as a linear combination of the $v_j$.
Also, define the diagonal matrix $\delta\Lambda = \diag(\delta\tilde\lambda_1, \delta\tilde\lambda_2, \ldots, \delta\tilde\lambda_m)$.
Then \cref{Seq:perteigdeg} can be rewritten as
\begin{linenomath}
\begin{align}
    W^\dagger X VV' &= W^\dagger VV' \delta\Lambda \notag \\
    (W^\dagger V)^{-1} W^\dagger X V &= V' \delta\Lambda V'{}^{-1} \label{Seq:perteigdiagdeg}
\end{align}
\end{linenomath}
Thus, the problem becomes diagonalizing the matrix on the left-hand side of \cref{Seq:perteigdiagdeg},
and the resultant eigenvalues will be the perturbations to the original degenerate eigenvalue $\tilde\lambda$.

When $\tilde M$ is Hermitian, we can choose $W$ and $V$ to be equal unitary matrices and
\cref{Seq:perteigdiagdeg} becomes the first-order perturbation theory result from quantum mechanics.

Let's now study the distribution of the eigenvalue perturbations $\delta\tilde\lambda_i$.
If we uniformly choose from $\{\delta\tilde\lambda_1, \delta\tilde\lambda_2, \ldots, \delta\tilde\lambda_m\}$ at random,\footnote{
This is the most reasonable thing to do when asking the question ``What's the perturbation
to eigenvalue $\tilde\lambda$?'' when the eigenvalue is degenerate.
Any other interpretation such as asking about the smallest value among the eigenvalue perturbations
would break the symmetry of the degeneracy.
} then the expected value is
\[
\frac{1}{m}\mathbb E[\tr(\delta\Lambda)] = \frac{1}{m}\mathbb E[\tr((W^\dagger V)^{-1} W^\dagger X V)] = 0.
\]
The last step used the fact that
every entry of $(W^\dagger V)^{-1} W^\dagger X V$ is a linear combination of entries of $X$,
all of which have zero mean.

Now, consider the case where all singular values of $F$ are distinct and positive. This is true almost surely
when $F$ is chosen to be the MaxEnt solution with random perturbations added to its entries to model
deviations from the MaxEnt solution. Assume also that the entries of $\eta$ are constant. WLOG, let them all be 1.
Then $\tilde M = \begin{bmatrix} 0 & -\alpha F \\ \beta F^\top & 0 \end{bmatrix}$.
According to \cref{Ssec:spectrum},
the nonzero eigenvalues are the $\pm$ square roots of the nonzero eigenvalues of $-\alpha\beta FF^\top$
(or, equivalently, $-\alpha\beta F^\top F$). Therefore, there are $2\min(S_R, S_C)$ nonzero eigenvalues
of the form $\pm i\sqrt{\alpha\beta}\sigma_k$ where $\sigma_k$ are the $\min(S_R, S_C)$ singular values of $F$,
and the remaining $|S_R - S_C|$ eigenvalues are zero.

WLOG, suppose $S_R \geq S_C$. Write the SVD of $F$ as $F = \sum_{k=1}^{S_C} \sigma_k v_k u_k^\top$.
By \cref{Ssec:spectrum}, the eigenvectors of $\tilde M$ corresponding to singular value $\sigma_k$ of $F$ are
$(\pm i\sqrt{\alpha\beta}\sigma_k v_k) \oplus \beta F^\top v_k = (\pm i\sqrt{\alpha\beta}\sigma_k v_k) \oplus (\beta\sigma_k u_k)$.
Let's scale it to $\left(\pm i\sqrt{\alpha}v_k\right) \oplus \left(\sqrt{\beta}u_k\right)$ for simplcity.
Similarly, we can obtain the corresponding left eigenvectors:
$\left(\mp i\sqrt{\beta}v_k^\top\right) \oplus \left(\sqrt{\alpha}u_k^\top\right)$.
Finally, if $v_{S_C+1}, \ldots, v_{S_R}$ is a basis of the nullspace of $F^\top$,
then $v_k \oplus 0$ ($S_C + 1 \leq k \leq S_R$) form a basis of the nullspace of $\tilde M$.

Consider the orthonormal basis consisting of the vectors $v_k \oplus 0$ ($1 \leq k \leq S_R$)
and $0 \oplus u_k$ ($1 \leq k \leq S_C$). In this basis, the eigenspaces of $\tilde M$ are solely a function of
$\alpha$ and $\beta$: the dependence on the entries of $F$ has been absorbed into the change of basis.
Therefore, if we always work in this basis constructed from the left and right singular vectors of $F$,
then the set of $V$ and $W$ matrices for the eigenspaces and left eigenspaces of $\tilde M$ will be independent of $F$.
Thus, the distribution of eigenvalue perturbations given by \cref{Seq:perteigdiagdeg} will be independent of $F$
as long as the distribution of the perturbation matrix $X$ is invariant under the change of basis.
Indeed, since the new basis is orthonormal, the change-of-basis matrix $B$ is orthogonal.
Then $(B^\top X B)_{i,j} = \sum_{k,l} B_{ki} B_{lj} X_{kl}$
is distributed normally with mean 0 and
\begin{linenomath}
\begin{align*}
    \cov((B^\top X B)_{i,j}, (B^\top X B)_{i',j'}) &= \sum_{k,l,k',l'} B_{ki} B_{lj} B_{k'i'} B_{l'j'} \cov(X_{kl}, X_{k'l'}) \\
    &= \sum_{k,l} B_{ki} B_{lj} B_{ki'} B_{lj'} \sigma^2 \\
    &= \delta_{ii'} \delta_{jj'} \sigma^2.
\end{align*}
\end{linenomath}
So $B^\top X B$ is a random matrix with i.i.d.\ normal random variables with mean 0 and variance $\sigma^2$, just like $X$.
This completes the proof that the first-order perturbations to the eigenvalues of $\tilde M$ are
independent of $F$ when the entries of $\eta$ are equal and the singular values of $F$ are distinct and positive.

\subsection{Perfectly-nested non-singular $F$: $F^\top F$ has larger eigenvalues that are less spread out when $N_p$ is high}

Let's treat off-diagonal elements of $F^\top F$ as a perturbation $\delta A$ to the diagonal part $A$ of $F^\top F$.
Note that both $A$ and $\delta A$ are symmetric matrices, so that we can directly cite perturbation theory
results from quantum mechanics, which apply to Hermitian matrices.

The eigenvalues of $A$ are just the elements on the diagonal, $a_{ii} = \vec f_i \cdot \vec f_i$.
When $N_p$ is high, the variance of the components of $f_i$ is high, i.e., the mean of the squares of
the components of $f_i$ is high (the square of the mean of the components is a constant $1/S_R^2$).
Thus, $\vec f_i \cdot \vec f_i$ is high, thus showing that the eigenvalues of $A$ are larger for larger $N_p$.

Let's now look how the perturbation $\delta A$ changes the eigenvalues.

For a non-degenerate eigenvalue $a_{ii}$ corresponding to the unit eigenvector $e_i$,
the first-order perturbation to the eigenvalue is $\delta A_{ii}$ and the second-order perturbation is
\begin{equation}
    \sum_{j \neq i} \frac{\delta A_{ij}^2}{a_{ii} - a_{jj}}.
\end{equation}
Each individual term in the sum is positive when $a_{ii} > a_{jj}$ and negative otherwise,
meaning that eigenvalues smaller than $a_{ii}$ push it upwards and eigenvalues larger than $a_{ii}$ push it downwards,
a phenomenon known as \textit{level repulsion} in quantum mechanics \citep{zwiebach2022quantum}.
The effect of level repulsion spreads
the eigenvalues out, with the effect greater when the perturbation is larger, i.e., when $N_p$ is smaller.

For a degenerate or near-degenerate eigenspace spanned by unit eigenvectors $e_i$ for $i \in I$, computing the second-order
perturbation according to the formula above gives values that explode. This is interpreted as
a nonzero first-order perturbation, which can be calculated using degenerate perturbation theory.
The result is that the first order perturbations to $a_{ii}$ for $i \in I$ are given by the eigenvalues of the symmetric matrix
$\widetilde{\delta A} = (\delta A_{ij})_{i,j\in I}$. The sum of these eigenvalues is $\tr(\widetilde{\delta A}) = 0$,
whereas the sum of the squares of the eigenvalues is
$\tr(\widetilde{\delta A}{}^2) = \sum_{i,j\in I} \delta A_{ij}^2$.
Thus, the original degenerate eigenvalues $a_{ii}$ ($i \in I$) of $A$ are now split
(known as \textit{level splitting} in quantum mechanics), and the spread of the
resulting eigenvalues is greater when the entries of $\widetilde{\delta A}$ are greater, i.e., when $N_p$ is smaller.

\subsection{Case where some consumers share the same set of resources}

Since the eigenvalues of $F^\top F$ are the squares of the singular values
of $F$, let's study those instead. Treating the random deviation from the MaxEnt matrix as a perturbation, i.e.,
$F = \overline F + \Delta F$, the zeros among the singular values of $\overline F$
become the singular values of $\widetilde{\Delta F}$, defined as $\Delta F$ with its domain restricted to
the nullspace of $\overline F$ and image projected onto the left nullspace of $\overline F$ (see below).
This is an $(S_R - S) \times (S_C - S)$ random matrix where $S$ is the rank of $F$.
If the entries of this random matrix are i.i.d.\ normal
with variance $\sigma^2$, then according to the Marchenko-Pastur distribution \citep{marchenko1967},
the smallest singular value is approximately
\begin{equation}
    \sigma\left|\sqrt{S_R - S} - \sqrt{S_C - S}\right| = \sigma\frac{|S_R - S_C|}{\sqrt{S_R - S} + \sqrt{S_C - S}}
\end{equation}
for large $S_R-S, S_C-S$.
The smallest singular value of $\widetilde{\Delta F}$ is larger
when $S$ is larger, i.e., when the topology is more nested according to $N_p$.
In this case, the smallest singular value of $F$ is also larger, translating to the smallest eigenvalues
of $F^\top F$ being larger, hence increasing stability.

\paragraph{Derivation of the perturbation to the zero singular values of a matrix}

Write the SVD of $\overline F$ as $\overline F = \sum_k \sigma_k U_k V_k^*$ where $\sigma_k$
are the \textit{distinct} singular values with multiplicities $d_k$, $U = \begin{bmatrix} U_1 & U_2 & \ldots U_k \end{bmatrix}$
is unitary, and $V = \begin{bmatrix} V_1 & V_2 & \ldots V_k \end{bmatrix}$ is unitary.
Write the SVD of the perturbed matrix $F$ as
\begin{equation} \label{Seq:SVD}
    \overline F + \Delta F = \sum_k (U_k + \Delta U_k)\tilde U_k(\sigma_k I_{d_k} + \Delta\Sigma_k)\tilde V_k^*(V_k + \Delta V_k)^*,
\end{equation}
where $\Delta\Sigma_k$ is diagonal, $\tilde U_k, \tilde V_k \in \mathrm{U}(d_k)$,
$\tilde U_k \tilde V_k^* \approx I_{d_k}$ to first order (unless $\sigma_k = 0$), $U_k^* \Delta U_k \approx 0$ to first order
and $V_k^* \Delta V_k \approx 0$ to first order. To first order, \cref{Seq:SVD} can be rewritten as
\begin{equation} \label{Seq:SVDexpand}
    \Delta F \approx \sum_k \left(\sigma_k (\Delta U_k) V_k^* + \sigma_k U_k (\Delta V_k)^* + \sigma_k U_k (\tilde U_k \tilde V_k^* - I_{d_k}) V_k^* + U_k \tilde U_k (\Delta\Sigma_k) \tilde V_k^* V_k^*\right).
\end{equation}
To obtain the singular value perturbations $\Delta\Sigma_{k_0}$,
we multiply from the left by $U_{k_0}^*$ and from the right by $V_{k_0}$ to obtain
\begin{equation}
    U_{k_0}^* (\Delta F) V_{k_0} \approx \sigma_{k_0} (\tilde U_{k_0} \tilde V_{k_0}^* - I_{d_{k_0}}) + \tilde U_{k_0} (\Delta\Sigma_k) \tilde V_{k_0}^*.
\end{equation}
If $\sigma_{k_0} = 0$, then this becomes
\begin{equation}
    U_{k_0}^* (\Delta F) V_{k_0} \approx \tilde U_{k_0} (\Delta\Sigma_k) \tilde V_{k_0}^*,
\end{equation}
as desired.

\section{Random topology sampling}

Traditionally, a random topology is generated
by setting each element of the adjacency matrix to an independent Bernoulli random variable. However,
this results in poor diversity in nestedness values.\footnote{
We generated $10^5$ adjacency matrices this way for $S_R = 6, S_C = 4, p = 0.5$, and only 1.4\% were perfectly nested.
This fraction decreases exponentially with $S_R$ and $S_C$, so that when $S_R = 8, S_C = 6$, we found that none
of the $10^5$ randomly generated adjacency matrices were perfectly nested.
} Instead, we use the following algorithm that allows us to freely adjust the nestedness through
the ``number of swaps'' parameter $s$:
\begin{enumerate}
    \item First, randomly generate a perfectly nested topology according to the procedure outlined below.
    \item We repeat $s$ times: uniformly choose an edge at random and move it to a uniformly random
    new location that doesn't have an edge. Equivalently, a uniformly chosen 1 and uniformly chosen
    0 in the adjacency matrix are swapped.
    \item Check whether or not the resultant network is valid, i.e., every species has positive degree
    and an energy flow solution exists under our assumption of energy equivalence. If the network
    is invalid, we start over.
\end{enumerate}
$s$ can be understood as a temperature parameter: when $s = 0$, the output network is perfectly nested;
as $s$ is increased, more swaps have been introduced to the initial perfectly nested network,
so that the output network is less nested.

We generated 25 random topologies for each value of $s$ between 0 and $0.4S_RS_C$.

\paragraph{Random nested topology generation}
A nested topology is defined by its linkage existence boundary.
It is made of several alternating vertical and horizontal segments from the top-right corner to the bottom-left corner
(e.g., red dashed lines in main text fig.~3 for $E'$ and $\tilde E$).
Let $S$ be the number of vertical segments (or, equivalently, horizontal segments);
this number is 6 in the left panel of fig.~3 for $E'$ and 4 in the right panel for $\tilde E$.
To generate a random nested topology, we first choose $S$ uniformly at random between 2 and $\min(S_R, S_C)$.
Afterwards, we choose the horizontal positions of the $S-1$ vertical segments (excluding the rightmost one)
uniformly at random (without replacement) from the $S_C - 1$ possible horizontal positions.
Similarly, we choose the vertical positions of the $S-1$ horizontal segments (excluding the bottommost one)
uniformly at random (without replacement) from the $S_R - 1$ possible vertical positions.

\section{Comparison with Existing Nestedness Metrics}

Here, we study the correlation between nestedness and instability using the following 4 metrics of nestedness
and 2 metrics of instability.

\textit{4 metrics of nestedness:}
\begin{itemize}
    \item The \textit{matrix dipole moment} $N_p$, as defined in the main text.
    \item \textit{Nestedness by overlap and decreasing fill} NODF \citep{almeida2008nodf}.
    \item The nestedness metric $N_T = 1 - T/100$, where $T$ is the \textit{matrix temperature}. 
    $T$ was defined first in 
    \citep{atmar1993matrixtemp},
    but the definition was ambiguous,
    so we implement $T$ based on BINMATNEST \citep{rodriguez2006binmatnest}.
    \item The nestedness metric $N_d = 1 - 4d/S_R S_C$, where $d$ is the \textit{nestedness discrepancy} \citep{brualdi1999nesteddiscrepancy}. Equivalently, this is $N_d = 1 - d_0$ where $d_0$
    is a normalized version of discrepancy introduced by 
    \citep{greve2006discrepancy0}.
\end{itemize}
To reduce computational costs, the algorithm for finding the maximally packed configuration is
by simply ordering the species in decreasing degree.

\textit{2 metrics of instability:}
\begin{itemize}
    \item $\overline m$, as defined in the main text.
    \item $p_{m > 0}$, the proportion of $m$ values that are positive among the $\tilde M$
    matrices samples for a given topology.
\end{itemize}

\begin{figure}[ht]
    \centering
    \includegraphics[width=\textwidth]{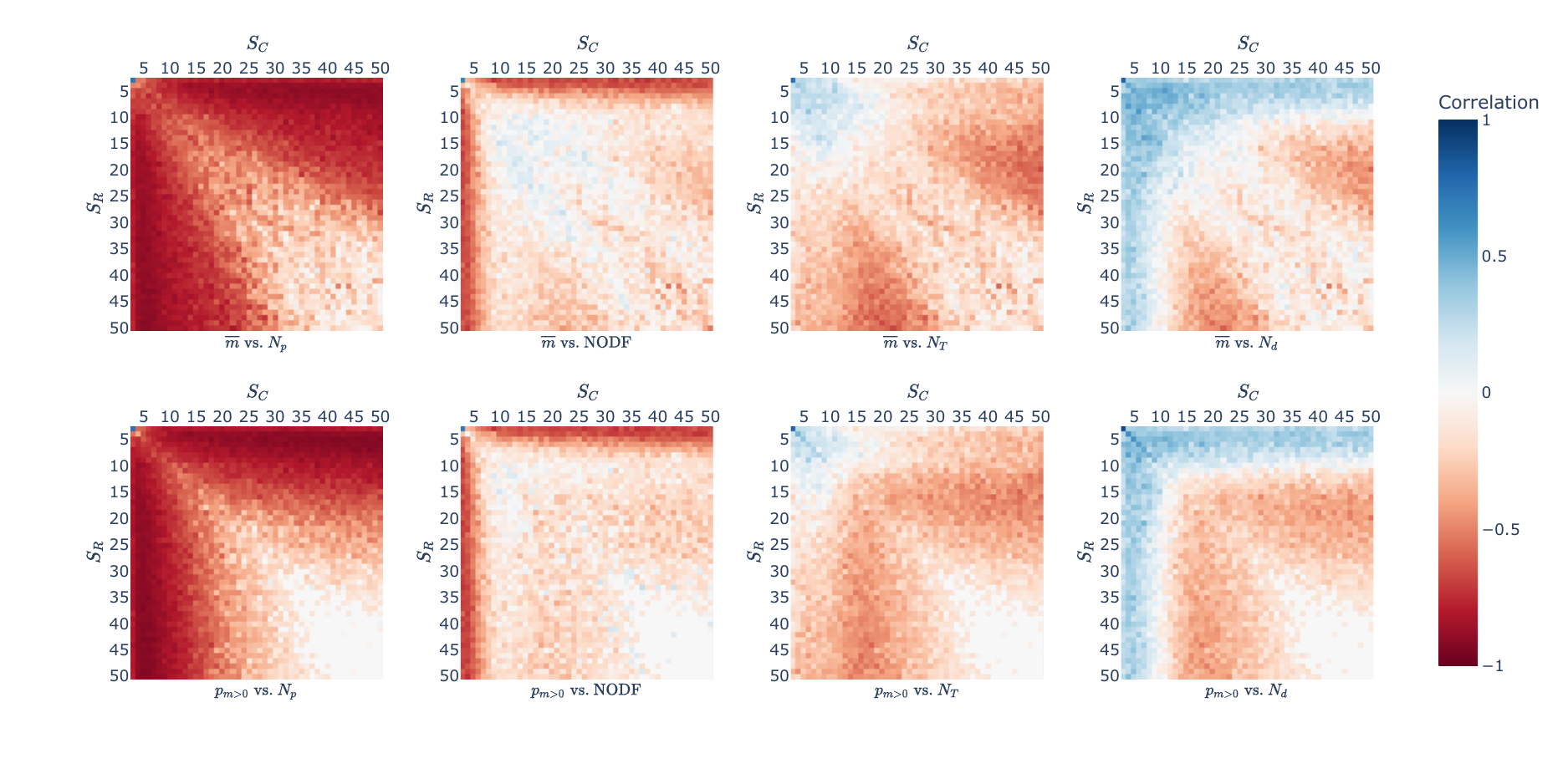}
    \caption{Correlation betwen instability and nestedness for each $(S_R, S_C)$ pair
    with $3 \leq S_R, S_C \leq 50$ for each instability metric ($\overline m, p_{m > 0}$)
    for each nestedness metric ($N_p, \text{NODF}, N_T, N_d$).}
    \label{Sfig:corrheatmaps}
\end{figure}

For each pair $S_R, S_C$ with $3 \leq S_R, S_C \leq 50$, we sampled 250 topologies
with a roughly equal number of topologies for each $s$ from 0 up to $0.4S_R S_C$.
For each topology, 250 $\tilde M$ matrices were sampled, from which the instability metrics
$\overline m$ and $p_{m > 0}$ were calculated. A linear regression was performed between
each of the two instability metrics and each of the four nestedness metrics, and the resultant
$r$-values are plotted on a heat map for each $(S_R, S_C)$ pair (\cref{Sfig:corrheatmaps}).

We notice significant negative correlation regardless of the instability metric or nestedness metric used,
as long as $S_R$ and $S_C$ are not too close to each other and are sufficiently large.
In addition, we notice that the negative correlation is present even when $S_R$ or $S_C$
are small if we use the $N_p$ or NODF metrics, as long as $(S_R, S_C) \neq (3, 3)$.
As expected, the negative correlations are the strongest for $N_p$.

\newpage

\section*{Supplementary Tables}

\begin{table}[h!]
\begin{spacing}{1}
\caption{Information regarding the food web networks involved in the empirical data analysis. For the ``Resources'' and ``Consumers'' columns, the format for each cell is [\# of species]: [list of species separated by commas].}
\end{spacing}
\label{tab:datasources}
\centering
\scriptsize
\begin{spacing}{1}
\begin{tabu}{p{0.5in}|[2pt]p{0.75in}|p{1.75in}|p{1.5in}|p{0.5in}|p{0.5in}}
Reference & Ecosystem type & Resources & Consumers & Total \# linkages & Topologically nested? \\ \tabucline[2pt]{-}
\citep{cross2011glencanyon} & Riparian & 6: Amorphous detritus, Diatoms, Leaves, Filamentous algae, Macrophytes, Fungi & 8: Potamopyrgus antipodarum, Gammarus lacustris, Tubificida (a), Physidae, Lumbricidae, Chironomidae, Simuliidae, Other $^*$ & 40 & Yes \\ \hline
\citep{bumpers2017foreststreams} & Forest streams & 14: Elmidae, Copepoda, Ceratopogonidae, Hexatoma, Diptera (non-Tanypodinae), Tanypodinae, Maccaffertium, Paraleptophlebia, Serratella, Amphinemura, Isoperla, Leuctra, Tallaperla, Wormaldia & 2: Desmognathus quadramaculatus, Eurycea wilderae & 28 & Yes \\ \hline
\citep{moriniere2003coralreef} & Mangrove; seagrass bed & 12: Tanaidacea, Copepoda, Isopoda, Amphipoda, Mysidecea, Bivalvia, Gastropoda, Decapoda, Fish, Sediment, Other $^\dagger$, Unidentified & 5: Haemulon flavolineatum, Haemulon sciurus, Lutjanus apodus, Lutjanus griseus, Ocyurus chrysurus & 41 & No \\ \tabucline{2-}
 & Coral reef & 4: Decapoda, Fish, Sediment, Unidentified &  & 13 & No \\ \hline
\citep{rudnick2005crustacea} & Freshwater tributary & 4: Detritus, Algae, Invertebrates, Inorganic matter & 2: Chinese mitten crab, Red swamp crayfish & 8 & Yes \\ \tabucline{2-}
\citep{judas1992earthworm} & Beech wood & 2: Inorganic matter, Non-particulate organic matter & 6: L. terrestris, L. castaneus, Aporrectodea caliginosa, Aporrectodea rosea, O. lacteum, O. cyaneum & 12 & Yes \\ \hline
\citep{baird1989chesapeake} & Estuary (mesohaline) & 2: Bay anchovy, Menhaden & 3: Blue fish, Summer flounder, Striped bass & 6 & Yes \\ \hline
\citep{goldwasser1993caribbean} & Tropical island & 2: Coleoptera adult, Adult spider & 2: Hummingbirds, Elaenia & 4 & Yes \\ \tabucline{3-}
 &  & 3: Coleoptera larva, Diptera larva, Ants & 4: Bullfinch, Coleoptera adult, Centipede, Orthoptera & 11 & Yes \\ \tabucline{3-}
 &  & 2: Collembola, Mites & 4: Coleoptera larva, Diptera larva, Ants, Thysanoptera & 8 & Yes \\ \tabucline{3-}
 &  & 2: Detritus, Fungi & 2: Collembola, Mites & 4 & Yes \\ \tabucline{3-}
 &  & 2: Wood, Detritus & 2: Coleoptera larva, Isoptera & 4 & Yes \\ \hline
\citep{christian1999seagrass} & Halodule wrightii community & 3: Meiofauna, Benthic algae, Detritus & 4: Brittle stars, Deposit-feeding peracaridan crustaceans, Deposit-feeding gastropods, Deposit-feeding polychaetes & 11 & Yes \\ \tabucline{3-}
 &  & 6: Meiofauna, Epiphyte-grazing amphipods, Deposit-feeding peracaridan crustaceans, Zooplankton, Macro-epiphytes, Benthic algae, Detritus & 4: Sheepshead minnow, Killifishes, Gobies and blennies, Pipefish and seahorses & 20 & No
\end{tabu}
\bigskip{}
\\
{\footnotesize
$^*$ Ostracoda, Nematoda, Sphaeridae, Cladocera, Copepoda, Tubificida (b) \\
$^\dagger$ Oligochaeta, Polychaeta, Echinoidea, Ostracoda, Seagrass, Foraminifera, Filamentous algae, Calcareous algae, Macroalgae
}
\end{spacing}
\end{table}

\end{document}